\documentstyle[prl,twocolumn,aps,floats,epsfig]{revtex}

\begin{document}

\draft
\twocolumn[\hsize\textwidth\columnwidth\hsize\csname
@twocolumnfalse\endcsname

\title{Anomaly in the Tunneling $I(V)$ Characteristics of 
Bi$_2$Sr$_2$CaCu$_2$O$_{8+x}$} 

\author{A. Mourachkine} 

\address{Free University of Brussels, B-1050 Brussels, Belgium} 

\date{{\bf JETP Lett. 77, 666 (2003)}}
\maketitle

\begin{abstract} 

Tunneling measurements have been carried out on slightly overdoped 
Bi$_2$Sr$_2$CaCu$_2$O$_{8+x}$ single crystals below and above the 
critical temperature by break-junctions and in-plane point-contacts. 
An anomaly was found in the tunneling $I(V)$ characteristics. Analysis 
of the data shows that the anomaly is caused by the superconducting 
condensate. In the extracted $I(V)$ characteristics of the condensate, 
the constant asymptotics points to the presence of one-dimensionality 
in Bi2212. The anomaly found here puts additional constraints on the 
final theory of high-$T_c$  superconductivity.

\end{abstract}

]

Soon after the discovery of superconductivity (SC) in cuprates \cite{Muller}, 
it became clear that the concept of the Fermi liquid is not applicable 
to the cuprates: the normal state properties of cuprates are markedly 
different from those of conventional metals \cite{Orenstein}. The pseudogap 
which manifests itself in electronic excitation spectra of cuprates above 
the critical temperature $T_c$, is one of the main features of 
high-$T_c$ SCs. There is a consensus on doping dependence of the pseudogap 
in hole-doped cuprates: the magnitude of the pseudogap decreases as the hole 
concentration increases. 
Angle-resolved photoemission spectroscopy (ARPES) measurements 
performed in Bi$_2$Sr$_2$CaCu$_2$O$_{8+x}$ (Bi2212) show that the 
ARPES spectra consist of two {\em independent} contributions---from the 
pseudogap (hump) and the SC condensate (quasiparticle peak) \cite{Fedorov}. 
As the temperature decreases, the quasiparticle peak appears in the spectra
slightly above $T_c$ on one side of the hump, meaning that the pseudogap and 
the SC gap coexist below $T_c$ in Bi2212. 

In addition to their peculiar normal-state properties, a number of 
experiments show that some SC properties of cuprates deviate from 
predictions of the BCS theory for conventional SCs \cite{Orenstein}. For 
example, the BCS isotope effect is almost absent in optimally doped 
cuprates. As another example, let us compare tunneling data obtained in 
cuprates and theoretical predictions for conventional SCs. 
Figure 1 shows a theoretical $I(V)$ curve for classical SCs (Fig. 6 in 
Ref. \cite{Tinkham}) and a tunneling $I(V)$ characteristic obtained in an 
underdoped Bi2212 single crystal (Fig. 1 in Ref.\cite{Miyakawa}). In the 
{\em tunneling}  regime, depending on the normal resistance of a junction, the 
theoretical $I(V)$ curve at high positive (low negative) bias lies somewhat 
below (above) the normal-state curve, as shown in Fig. 1a. In  conventional 
SCs, the so-called Blonder-Tinkham-Klapwijk (BTK) predictions are verified 
by tunneling experiments. In cuprates, as one can see in Fig. 1b, the  BTK 
theory is violated: the $I(V)$ curve at high  positive (low negative) bias 
passes not below (above) the line which is parallel to the $I(V)$ curve at 
high bias but far above (below) the line. This fact cannot be explained by 
the d-wave symmetry of the order  parameter. This question, for the first 
time, was raised elsewhere \cite{AMour1}. This finding is the main 
motivation of the present work. 
\begin{figure} 
\leftskip-10pt
\epsfxsize=0.9\columnwidth
\centerline{\epsffile{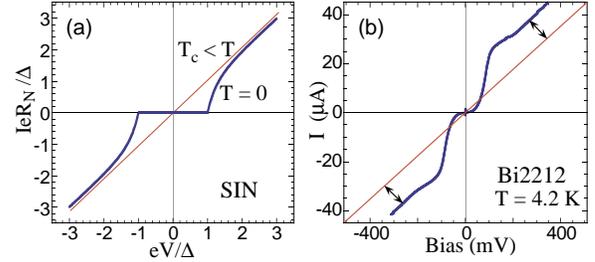}}
\vspace{2mm} 
\caption{(a) Theoretical tunneling $I(V)$ characteristic at 
$T$ = 0 for a SIN junction of a SC with the isotropic energy gap 
\protect\cite{Tinkham}. The line is the normal-state curve. 
(b) Measured $I(V)$ curve in a SIS junction of an 
underdoped Bi2212 single crystal with $T_c$ = 83 K, obtained at 
$T$ = 4.2 K \protect\cite{Miyakawa}. The line is parallel to the $I(V)$ 
curve at high bias, and the arrows show the offset from the line.}
\label{fig1}
\end{figure} 

The data shown in Fig. 1b are obtained in an underdoped Bi2212 single 
crystal in a SC-insulator-SC (SIS) tunneling junction.
If the anomaly is an intrinsic feature of SC in Bi2212, it has to be present 
in tunneling spectra in the overdoped region of Bi2212 as well. Second, if
it is not a SIS-junction effect, it must also manifest itself in
SC-insulator-normal metal (SIN) junctions. Third, the line  in Fig. 1b is 
not the normal-state curve, therefore it is necessary to obtain tunneling  
spectra in the normal state. Finally, knowing the normal-state curves one 
can estimate the contribution in tunneling spectra from the SC condensate. 
This work deals with the questions raised above. Tunneling measurements 
presented in this paper have been performed on overdoped Bi2212 single 
crystals below and above $T_c$ by break-junctions and in-plane 
point-contacts, 
which reveal that the anomaly found in the tunneling $I(V)$ characteristics 
originates from the SC condensate. In the extracted $I(V)$ characteristics 
of the SC condensate, the constant asymptotics points to the presence of 
one-dimensionality in Bi2212. The anomaly found here puts additional 
constraints on the final theory of SC in the cuprates. 

It is important noting that the anomaly in tunneling $I(V)$ curves was 
already discussed in Ref. \cite{AMour1}; however, in Ref. \cite{AMour1}, the 
anomaly was inferred from abnormally-looking tunneling characteristics. In 
this work, we show that this anomaly is intrinsically present in {\em any} 
$I(V)$ curve obtained below $T_c$ in Bi2212. Also, the $I(V)$ 
characteristics measured above $T_c$ were {\em not} considered in 
Ref. \cite{AMour1}. \begin{figure} 
\leftskip-10pt
\epsfxsize=1.0\columnwidth
\centerline{\epsffile{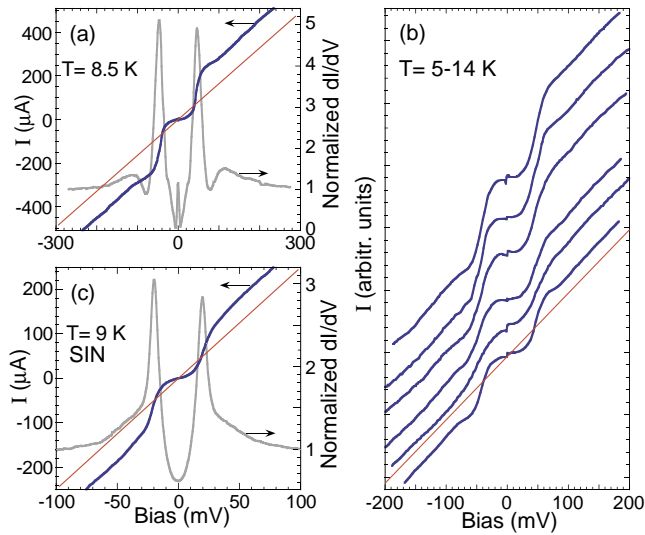}}
\vspace{2mm} 
\caption{(a) Tunneling $I(V)$ and $dI(V)/dV$ characteristics obtained 
at $T$ = 8.5 K in a SIS junction of an overdoped Bi2212 single crystal with 
$T_c$ = 88 K. 
(b) Set of $I(V)$ curves obtained at $T$ = 5--14 K in different overdoped 
Bi2212 single crystals with $T_c$ = 87--89 K. The curves are offset for 
clarity. 
(c) $I(V)$ and $dI(V)/dV$ characteristics obtained at $T$ = 9 K in a SIN 
junction of an overdoped Bi2212 single crystal with $T_c$ = 87.5 K. 
In all plots, the lines are parallel to the $I(V)$ curves at high bias. 
The label of the horizontal axis in plot (a) is the same as in plot (c).}
\label{fig2}
\end{figure} 

The overdoped Bi2212 single crystals were grown using the self-flux 
method as described elsewhere \cite{AMour2}. The $T_c$ value was 
determined by the four-contact method. The transition width is less than 
1 K. Experimental details of the measurement setup can be 
found elsewhere \cite{AMour2}. In short, many break-junctions were 
prepared by gluing a sample with epoxy on a flexible insulating 
substrate, and then were broken in the $ab$\,-plane by bending the
substrate with a  differential screw at low temperature in a He
ambient. The electrical contacts (typically with the resistance of a few 
ohms) are made by attaching gold wires to a crystal with silver paint. The 
$I(V)$ and $dI(V)/dV$ characteristics are determined by the four-terminal 
method by using the standard lock-in modulation technique. 
In in-plane SIN tunneling junctions, Pt-Ir wires sharpened mechanically 
were used as normal tips. 

Figure 2a shows SIS tunneling spectra obtained by a break-junction 
at $T$ = 8.5 K in an overdoped Bi2212 single crystal with 
critical temperature $T_c$ = 88 K. The conductance $dI(V)/dV$ exhibits 
typical features of SIS-junction conductance data in Bi2212: well-defined 
quasiparticle peaks, a zero-bias peak due to the Josephson current, dips and 
humps outside the gap structure. As the main result, the $I(V)$ characteristic 
in Fig. 2a is similar to the $I(V)$ curve shown in Fig. 1b. The gap magnitude 
$\Delta$, however, is smaller in the overdoped region. Figure 2b shows a 
set of tunneling $I(V)$ characteristics obtained in different overdoped 
Bi2212 single crystals with $T_c$ = 87--89 K. Comparing Figs. 1b, 2a 
and 2b it is evident that the anomaly is present not only in the underdoped 
region of Bi2212, but in the overdoped region as well. 
\begin{figure} 
\leftskip-10pt
\epsfxsize=1.0\columnwidth
\centerline{\epsffile{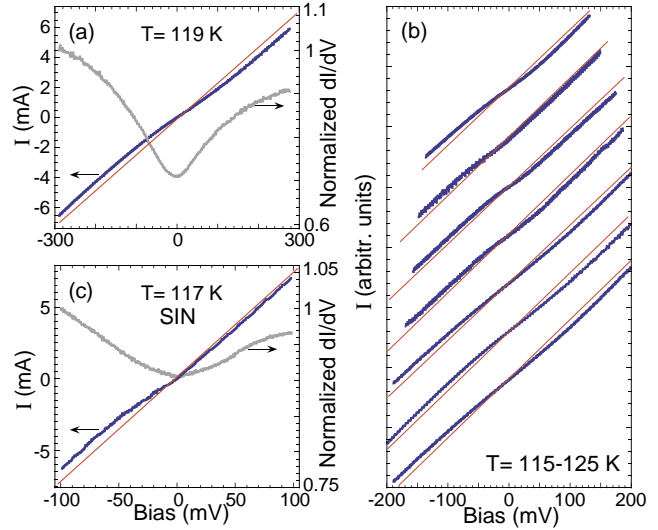}}
\vspace{2mm} 
\caption{(a) $I(V)$ and $dI(V)/dV$ characteristics obtained at 
$T$ = 119 K in the same SIS junction as those in Fig. 2a. 
(b) Set of $I(V)$ curves obtained at $T$ = 115--125 K in different overdoped 
Bi2212 single crystals with $T_{c}$ = 87--89 K. The curves are offset for 
clarity. 
(c) $I(V)$ and $dI(V)/dV$ characteristics obtained at $T$ = 117 K in the 
same SIN junction as those in Fig. 2c. In all plots, the lines are parallel to 
the $I(V)$ curves at high bias. 
The label of the horizontal axis in plot (a) is the same as in plot (c).}
\label{fig3}
\end{figure} 

Figure 2c depicts tunneling spectra which are similar to those in Fig. 2a, 
but obtained in a SIN junction. Because of the SIN junction, the quasiparticle 
peaks in the conductance shown in Fig. 2c appear at a bias twice smaller 
than the peak bias in Fig. 2a. The $I(V)$ curve in Fig. 2c clearly indicates 
that the observed anomaly is not a SIS-junction effect but an intrinsic 
feature of tunneling $I(V)$ characteristics obtained in Bi2212. 

The next question which we discuss is how to obtain tunneling $I(V)$ 
characteristics in the normal state. There are, at least, three different 
solutions. The first two consist in applying a magnetic field below $T_c$, 
while the third solution is to measure $I(V)$ above $T_c$. In the 
first case, a magnetic field with a magnitude larger than the upper critical 
magnetic field $H_{c2}$ in Bi2212 renders the whole sample normal. In the 
second case, by applying a magnetic field with a magnitude larger than the 
lower critical field $H_{c1}$, vortices will enter the sample. In the latter 
case, the normal-state characteristics can be obtained inside vortex cores. 
Because $H_{c2}$ in Bi2212 is very large, the first solution cannot be 
realized in laboratory conditions. The second solution seems to be suitable; 
however, tunneling spectra obtained inside vortex cores have subgap 
structures which are usually interpreted as a manifestation of bound states 
\cite{Pan}. Moreover, technically, it is not easy to realize such measurements. 
So, we are left with the straightforward solution: to measure $I(V)$ 
characteristics somewhat above $T_c$. The main disadvantage of 
measurements performed above $T_c$ is the presence of substantial 
thermal fluctuations. 
\begin{figure}[t]
\leftskip-10pt
\epsfxsize=1.0\columnwidth
\centerline{\epsffile{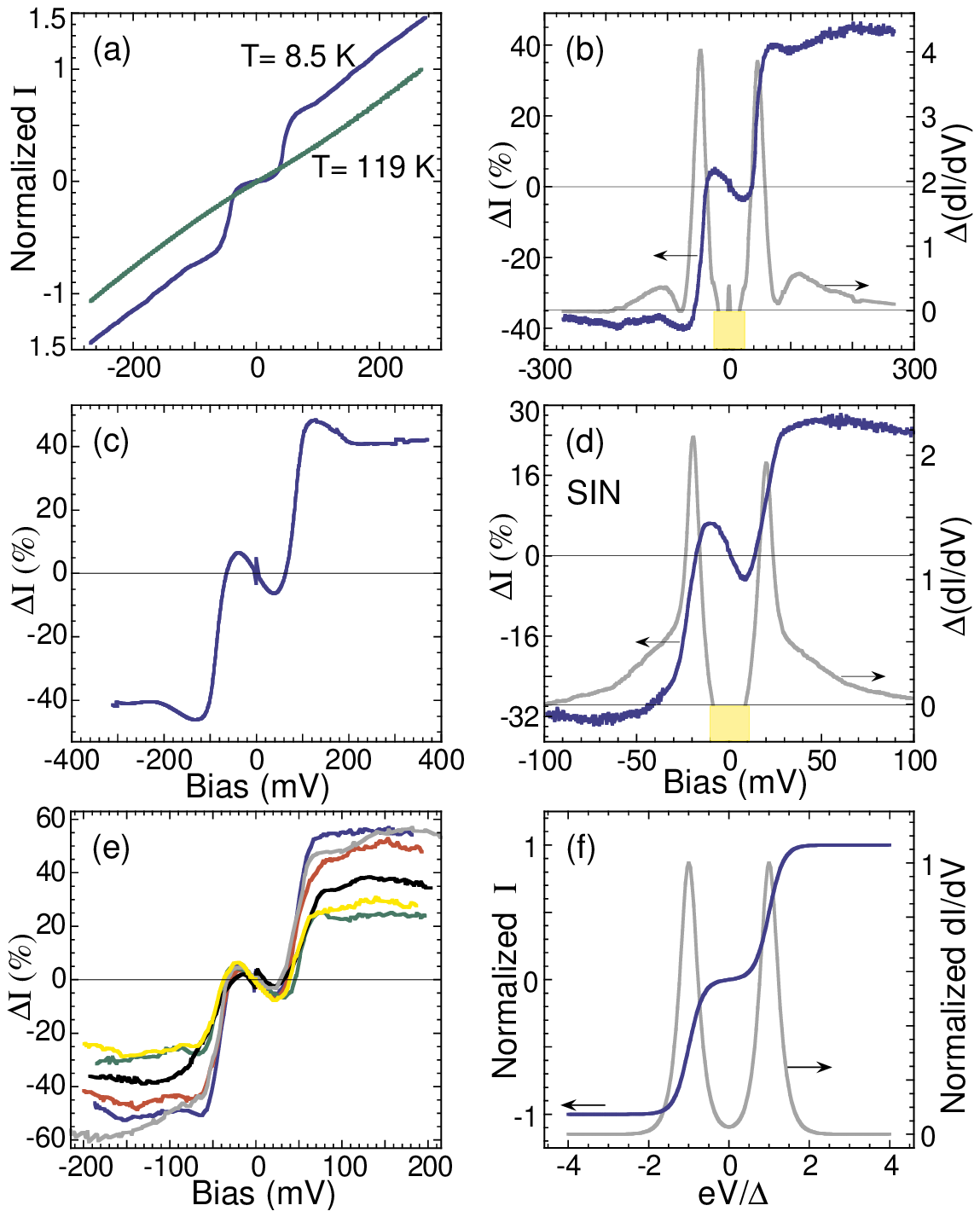}}
\vspace{2mm}
\label{fig4}
\end{figure} 
 
Figure 3a depicts tunneling spectra obtained at $T$ = 119 K in the same 
Bi2212 single crystal as those in Fig. 2a. Figure 3b presents a set of 
$I(V)$ curves measured at $T$ = 115--125 K in different overdoped Bi2212 
single crystals. The data in Figs. 3a and 3b are obtained in SIS junctions. 
Figure 3c shows tunneling data obtained in a SIN junction at $T$ = 117 K in 
the same Bi2212 single crystal as those in Fig. 2c. The temperatures 
between 115 and 125 K presented in Fig. 3 were not chosen by accident: in 
slightly overdoped Bi2212, the onset of SC occurs at 110--116 K 
\cite{AMour2}. This temperature range is above the onset of SC in our 
Bi2212 samples and, as a consequence, the data in Figs. 3a--3c mainly 
provide the contribution from the normal-state pseudogap. 
\begin{figure}[t] 
\caption{(a) Two normalized $I(V)$ characteristics from Figs. 2a and 3a,
obtained in the same Bi2212 single crystal ($T_c$ = 88 K) at different 
temperatures. The normalization procedure: the normal-state curve is 
normalized by its value at maximum positive bias, and the other curve is 
adjusted to be parallel at high bias to the normalized normal-state curve.   
(b) Difference between the two $I(V)$ curves presented in plot (a), and 
the difference between their normalized conductances shown in Figs. 2a 
and 3a. 
(c) Difference between the $I(V)$ curve and the line, shown in Fig. 1b, 
which were normalized before subtraction as those in plot (a). 
The straight line is used as an estimate of the normal-state curve. 
(d) Difference between the two $I(V)$ curves from Figs. 2c and 3c, 
normalized before subtraction as those in plot (a), and the difference 
between their normalized conductances. The data are obtained in a SIN 
junction. 
(e) Differences between $I(V)$ characteristics shown in Fig. 2b and 
their corresponding normal-state curves, normalized before 
subtraction as those in plot (a). 
(f) Idealized $I(V)$ characteristic of SC condensate in a SIN junction, and 
its first derivative. The curves are normalized by their maximum values.
In plots (b) and (d), the grey boxes cover the parts of the 
conductances, which are below zero and have no physical meaning.
In plots (b)--(e), the current difference is presented in \%. In plots (a) and 
(b), the labels of horizontal axes are the same as those in plots (c) and (d).}
\label{fig4}
\end{figure} 

Considering the $I(V)$ characteristics shown in Fig. 3, one can see that, at 
high positive (low negative) bias, they pass somewhat below 
(above) the line which is parallel to the $I(V)$ curves at high bias. Thus, 
the anomaly in the $I(V)$ characteristics, found below $T_{c}$, vanishes 
somewhat above $T_{c}$. Then, it is obvious that the anomaly in the $I(V)$ 
curves originates 
from the SC condensate. In Fig. 3, one can see that the straight line can be 
used as a first approximation for a normal-state $I(V)$ curve, for example, 
in Fig. 1b. In the $I(V)$ curves measured above $T_c$, a small ``negative'' 
offset from straight lines is caused by the pseudogap: measurements 
performed in underdoped Bi2212 show that this offset is larger than that 
in Fig. 3. Thus, it scales with magnitude of the pseudogap. 

We turn now to the last question raised above. From the tunneling 
characteristics obtained deep below $T_c$ and somewhat above $T_c$, 
by taking the difference between the spectra one can estimate a contribution 
in the tunneling spectra from the SC condensate. This is equivalent 
to the procedure usually used in neutron scattering measurements. In our 
case, however, this procedure only leads to an {\em estimation} of the 
contribution because by subtracting the spectra we assume that the 
pseudogap crosses $T_c$ without modification. This is not obvious, 
particularly, at low bias \cite{note}. 

In order to compare two sets of tunneling spectra, they have to be 
normalized. The conductance curves can be easily normalized at high bias. 
How to normalize the corresponding $I(V)$ curves is not a trivial question. 
The conductance curves at high bias, thus, far away from the gap structure, 
are almost constant. Consequently, in a SIS junction, by normalizing 
two conductance curves at high bias, the equation $(dI(V)/dV)_{1,norm} 
\simeq (dI(V)/dV)_{2,norm}$  holds at bias $|V| \gg 2 \Delta /e$, where 
$e$ is the  electron charge. By integrating the equation we have 
$I(V)_{1,norm} \simeq I(V)_{2,norm} + C$, where $C$ is the constant, 
meaning that the corresponding $I(V)_{i,norm}$ curves are parallel to each 
other at high bias.  

In Fig. 4a, the two $I(V)$ characteristics from Figs. 2a and 3a 
were normalized as described in the previous paragraph: the normal-state 
$I(V)$ curve is normalized by its value at maximum positive bias, 
and the $I(V)$ curve from Fig. 2a is adjusted to be parallel at high bias to 
the normalized normal-state curve. Figure 4b depicts their difference as 
well as the difference between their normalized 
conductances. The same procedure has been done for the tunneling spectra 
obtained in a SIN junction, which are shown in Figs. 2c and 3c. The 
differences are depicted in Fig. 4d. The conductances in Figs. 4b and 4d 
are presented as examples; however, they are not the primary focus of the 
study: we discuss exclusively the $I(V)$ characteristics.  The $I(V)$ data 
shown in Fig. 1b, which are obtained in an underdoped Bi2212 single crystal, 
went through the same normalization procedure, and the result is presented 
in Fig. 4c. In the latter case, the straight line shown in Fig. 1b was used 
as a normal-state curve. The $I(V)$ characteristics shown in Fig. 2b and 
their corresponding normal-state curves were normalized in the same 
manner, and the obtained differences are presented in Fig. 4e.

Analyzing the $I(V)$ characteristics shown in Figs. 4b--4e, which, in 
first approximation, represent the contribution from the SC condensate, it 
is easy to observe the general trends of the curves. First, at high bias, the 
curves reach a plateau value. Second, at the gap bias, the curves 
rise/fall sharply. Last, at low bias, the curves go to zero. In Figs. 4b--e, 
the negative slope of the curves at low bias, implying a negative differential 
resistance, is an artifact. This artifact is simply a consequence of the rough 
estimation used here \cite{note}. {\it Tout ensemble}, Figure 4f depicts an 
idealized $I(V)$ characteristic summarizing the observed tendencies. The 
first derivative of the $I(V)$ characteristic is also shown in Fig. 4f. 
The constant asymptotics of the $I(V)$ characteristic are the 
fingerprints of the presence of one-dimensionality in the system 
\cite{AMour1}. 
In Figs. 4b--4e, one can see that the contribution from the SC condensate 
in {\em these} $I(V)$ characteristics at high bias is 25--55 \% above 
the contribution from the pseudogap. It is {\em important} noting that the 
asymptotics of some $I(V)$ curves obtained in Bi2212 below $T_c$ look 
similar to those in Fig. 1a; {\em however}, the anomaly discussed here 
always appears in the difference obtained between two normalized $I(V)$ 
curves measured {\em below} and {\em above} $T_c$ in one sample. 

As shown elsewhere \cite{AMour1}, the $I(V)$ characteristic 
in Fig. 4f can be fitted by the hyperbolic function 
$f(V) = A \times (tanh[(eV - \Delta )/eV_0] + tanh[(eV + \Delta )/eV_0])$, 
where $e$ is the electron charge; $V$ is the bias; $\Delta$ is the maximum 
SC gap, and $A$ and $V_0$ are the constants. The conductance $dI(V)/dV$ 
curve can  be well fitted by the derivative $[f(V)]' = 
A_1 \times (sech^2[(eV - \Delta )/eV_0] + sech^2[(eV + \Delta )/eV_0])$. 
The hyperbolic $tanh$ and $sech^2$ fits are in good agreement with
predictions of the bisoliton model \cite{Davydov} which is based on the 
nonlinear Schr\"{o}dinger equation. 
The bisoliton theory utilizes the concept of {\em bisolitons}---electron
(or hole) pairs coupled in a singlet state due to local deformation of the
lattice \cite{Davydov}. 

Finally, it is worth mentioning that the anomaly found here is also present in 
an $I(V)$ characteristic obtained in optimally doped YBa$_2$C$_3$O$_{6.95}$ 
\cite{YBCO}. It is interesting that the anomaly is also present in the $I(V)$ 
characteristics measured in some {\em non}-superconducting 
materials---in the stripe-ordered perovskite 
La$_{1.4}$Sr$_{1.6}$Mn$_2$O$_7$ \cite{mang} and in 
quasi-one-dimensional charge-density-wave conductor NbSe$_3$ \cite{NbSe} 
(in NbSe$_3$, the anomaly is at zero bias). However, this issue is already 
beyond the scope of this study, and will be discussed elsewhere. 

In summary, tunneling measurements have been carried out on slightly 
overdoped Bi$_2$Sr$_2$CaCu$_2$O$_{8+x}$ single crystals below and 
above $T_c$ by break-junctions and point-contacts. An anomaly was found 
in the tunneling $I(V)$ characteristics. Analysis of the data shows that the 
anomaly is caused by the superconducting condensate. 
The constant asymptotics of the $I(V)$ characteristic of the condensate, 
shown in Fig. 4f, are the fingerprints of the presence of 
one-dimensionality in Bi2212 \cite{AMour1}. 
The anomaly found here puts additional constraints on the final theory of 
high-$T_c$ superconductivity. 

I would like to thank N. Miyakawa for sending the data from 
Ref.\cite{Miyakawa}, Yu. I. Latyshev for sending unpublished data 
related to Ref.\cite{NbSe}, and V. Z. Kresin for comments on the manuscript. 

\vspace*{-10mm}Ê


\begin{references} 
\vspace{-17mm} 

\bibitem{Muller} J. G. Bednorz and K. A. M\"{u}ller, Z. Phys. B {\bf 64}, 189 
(1986). 

\bibitem{Orenstein} J. Orenstein and A. J. Millis, Science {\bf 288}, 468 
(2000). 

\bibitem{Fedorov} A. V. Fedorov, T. Valla, P. D. Johnson, {\it et al.}, 
Phys. Rev. Lett. {\bf 82}, 2179 (1999). 

\bibitem{Tinkham} G. E. Blonder, M. Tinkham, and T. M. Klapwijk,
Phys. Rev. B {\bf 25}, 4515 (1982). 

\bibitem{Miyakawa} N. Miyakawa, P. Guptasarma, J. F. Zasadzinski, 
{\it et al}., Phys. Rev. Lett. {\bf 80}, 157 (1998). 

\bibitem{AMour1} A. Mourachkine, Europhys. Lett. {\bf 55}, 559 (2001). 

\bibitem{AMour2} A. Mourachkine, Europhys. Lett. {\bf 49}, 86 (2000). 

\bibitem{Pan} S. H. Pan, E. W. Hudson, A. K. Gupta, {\it et al.}, 
Phys. Rev. Lett. {\bf 85}, 1536 (2000). 

\bibitem{note} Crossing $T_c$ from above, the pseudogap undergoes a strong 
renormalization of quasiparticle excitations at low bias (see Fig. 3 in 
Ref. \cite{Fedorov}). 

\bibitem{Davydov} A. S. Davydov, Phys. Rep. {\bf 190}, 191 (1990). 

\bibitem{YBCO} A. G. Sun, A. Truscott, A. S. Katz, {\it et al.}, 
Phys. Rev. B {\bf 54}, 6734 (1996).

\bibitem{mang} T. Nachtrab, S. Heim, M. M\"ossle, {\it et al.}, 
Phys. Rev. B {\bf 65}, 012410 (2002). 

\bibitem{NbSe} Yu. I. Latyshev, A. A. Sinchenko, L. N. Bulaevskii, {\it et al.}, 
JETP Lett. {\bf 75}, 103 (2002).  


\end{references}
\end{document}